# Astro2020 Science White Paper
# The Nature of Low-Density Star Formation

**Thematic Areas:**

Resolved Stellar Populations and their Environments (Primary)
Galaxy Evolution (Secondary)

**Principal Author:**

**David Thilker**
Johns Hopkins University
dthilker@jhu.edu
+1-410-516-3861

**Co-authors:**

**Janice Lee** (Caltech/IPAC), Peter Capak, (Caltech/IPAC), David Cook (Caltech/IPAC),
Danny Dale (U. Wyoming), Bruce Elmegreen (IBM), Armando Gil de Paz (U. Madrid),
Jay Gallagher (U. Wisconsin), Deidre Hunter (Lowell Observatory),
Adam Leroy (Ohio State U.), Gerhardt Meurer (U. Western Australia - ICRAR),
D.J. Pisano (West Virginia U./GBO), Marc Rafelski (STScI/JHU),
Monica Tosi (INAF-Bologna), Aida Wofford (IA UNAM - Ensenada)

**Abstract**:

How do stars manage to form within low-density, HI-dominated gas? Such environments provide a laboratory for studying star formation with physical conditions distinct from starbursts and the metal-rich disks of spiral galaxies where most effort has been invested. Here we outline fundamental open questions about the nature of star formation at low-density. We describe the wide-field, high-resolution UV-optical-IR-radio observations of stars, star clusters, and gas clouds in nearby galaxies needed in the 2020's to provide definitive answers, essential for development of a complete theory of star formation.

**New Required Capabilities / Relevant Projects Emphasized:**

Ultra-wide-field (deg-scale), HST-resolution UV-optical imaging / CASTOR
Multi-object UV-opt. (R>300) spectroscopy at >HST sensitivity and resolution / LUVOIR, HabEx
High-collecting area, high surface brightness sensitivity interferometry at cm,mm $\lambda$'s / ngVLA
High survey speed FIR imaging spectroscopy at 2× Herschel resolution / Origins Space Telescope
Order of magnitude (+) leap in mm $\lambda$ spectral line mapping speed / large heterodyne arrays

# Charting a Path to Understanding Star Formation in All Conditions

Star formation (SF) is a principle driver of galaxy evolution. It occurs under an enormous range of conditions, as metallicity, gas richness, the interstellar radiation field and phase balance vary dramatically across cosmic space and time. Decades of observations across the electromagnetic spectrum have enabled the detailed characterization of SF in the environments of starbursts and the metal-rich disks of spiral galaxies.[0] We know far less about how stars manage to form within low-density, HI dominated gas.[1] Such environments are important because they characterize star forming dwarfs, the most common type of gas-rich galaxy and building blocks of larger systems, as well as the outskirts of more massive galaxies, the sites of cold gas reservoirs for inner disks, where galaxy growth occurs today, and where disks must end and meet the intergalactic medium. Presumably, it is within such environments that the first stars formed. Addressing this shortcoming is crucial for understanding key parameters that influence the SF and chemical enrichment that occurs in the vast majority of neutral HI gas in the universe out to high redshift, otherwise accessible only as absorption line systems such as DLAs.[3,4] Fundamentally, it is essential for development of complete theory of star formation.

GALEX[2] was a revolutionary probe of recent star formation in this low surface brightness (LSB) regime. Its wide (1.2°) field-of-view enabled discovery of "XUV disks" with UV-bright SF extending to several times beyond the optical disk (Fig. 1).[5,6,7] GALEX provided the UV data for studies that revealed more SF in low-density environments than previously inferred from H$\alpha$ observations.[8,9,10,11] This renewed interest in understanding the physics that sets the conversion efficiency of gas to stars in such environments [12,13,14] (top-right Fig. 1). It motivated development of a new generation of stellar population synthesis models required for inferring the physical properties of populations where the massive end of the stellar initial mass function is not fully sampled.[15,16] It leads to questions about the fraction of stars formed in bound clusters ($\Gamma$) as a function of density[17,18] (Fig. 1) and how variations in $\Gamma$ might mirror changing cloud populations.

Attention was thus focused on characterization of ensemble properties of SF in its primary units: young stars, clusters, and atomic/molecular clouds. That is, in low-density environments, do the mass functions, filling factors and spatial distributions of these quanta simply reflect ordinary activity proceeding at a slower, perhaps intermittent, pace? Or, for example, are mass functions intrinsically distinct from the Salpeter/Chabrier/Kroupa distributions for stars, or -2 power-law distribution for clusters and clouds?[19,20,21] Do hierarchical distributions of SF quanta at low-density reveal a record of fundamentally different physical processes dominating in this environment?

Obtaining the required observations of stars, star clusters, and gas clouds to provide definitive answers to these questions has been infeasible. Current facilities (e.g., HST, ALMA) are able to support such study of the inner disk[22,23,24,25,26] and starbursts in nearby galaxies.[27] However, to probe low-density environments, very large areas must be mapped in at least the UV, optical, and radio, to collect a statistical amount of SF for study (Fig. 2). For example, to obtain UV/optical resolved stellar population imaging for half of M83's archetypal XUV disk (0.5° x 0.25°, Fig. 1) is costly (~70 HST orbits). Yet, the integrated SFR over such a region is a mere ~ 0.02 $M_\odot$ $yr^{-1}$, and only 10-100 star clusters would be captured, subjecting any analysis to large Poisson uncertainties. Studying the ISM in these environments poses its own set of long-standing challenges. Stars should form from molecular clouds, but detection of the molecular tracers is impeded by low metallicity and strong UV radiation.[28] Observations of nearby galaxies which resolve gravitationally bound HI clouds at ~1" GMC scales that could host molecule and star formation is in principle possible with the VLA, but the time requirements are prohibitive.[29] Statistically sound investigations probing the full range of $\Sigma_{SFR}$ with stars and gas, in particular controlling for galactic environmental parameters such as metallicity, and mitigating against degeneracies between mass function shape and SF history, would require years of time with HST and the JVLA. Here, we outline critical problems that can be solved with new capabilities in the 2020's to achieve a more complete understanding of SF in all of its myriad forms.

***When SF is struggling to occur, how much SF activity is locked into long-lived clusters?***
Most stars are born in clustered environments (including associations), but their subsequent fate is far less clear. The cluster formation efficiency, $\Gamma = SFR_{cluster} / SFR_{total}$ (fraction of SF occurring in bound clusters[30]) at low $\Sigma_{SFR}$ is vigorously debated with evidence of reduced efficiency[18] (Fig. 1) but counterclaims of nearly constant $\Gamma$.[31] Additional constraint on the theoretical models[17] (predicting a downturn vs. $\Sigma_{gas}$ hence $\Sigma_{SFR}$) is now vital. The outer disks of galaxies are an ideal place to make the $\Gamma$ measurement, but a census of sufficient integrated SFR is necessary, implying a wide area. Total <SFR> of ~ few x $10^{-2}$ $M_\odot yr^{-1}$ over a 300-500 Myr duration is required per each independent sample of an environmental regime, otherwise the lowest $\Gamma$ values ($\Gamma < 0.1$) become indistinguishable due to low expected cluster number counts. Note that it is essential to exclude the youngest clusters, possibly dissolving until ~10 Myr, on the basis of UV-optical SED fitting. A $\Gamma$-$\Sigma_{SFR}$ study conducted at the extreme limit of SF activity, in an environment less hostile to dynamical cluster disruption, would also gauge the importance of internal vs. external cluster destruction mechanisms with clarity impossible in the complex inner disk and inter-arm zones.

***Are there upper IMF variations in LSB star forming environments? If so, what drives them?***
The stellar initial mass function (IMF) is a fundamental parameter that encodes the complex physics of SF, and is crucial for interpreting most observations in extragalactic astronomy. Since the idea of the "original" mass function was introduced by Salpeter in 1955, considerable effort has been invested to verify its form, and historically most studies have concluded that the IMF is invariant[32]. However, studies of LSB environments have suggested a possible deficiency of high mass stars. Forward-modeling of the luminosity function of main sequence stars[33,34,35] show trends in the IMF slope ($\alpha$) above $1 M_\odot$, or potentially severe limits on the maximum stellar mass ($M_u$), perhaps linked to the low pressures, gas densities, $\Sigma_{SFR}$ and/or metallicity [9,36,37] of outer disks. Indirect constraints from integrated light studies (e.g., M/L ratio, deficiencies of H$\alpha$ emission)[38,39,10,40,41,9] frequently support this picture, but there are exceptions[42]. Efforts in the next decade must address the significant limitations of current studies in order to provide conclusive results. IMF experiments must be conducted with large samples of massive stars from different galaxies to (1) probe a range of LSB environments (e.g. varied pressure, metallicity, local dynamics) and provide insight into the drivers of possible variations, (2) overcome Poisson uncertainties, (3) address degeneracies between IMF shape and star formation history (e.g., by averaging over many independent regions, and using relatively IMF-insensitive core He burning stars, best distinguished in blue/UV-optical CMDs, to directly derive the SFH in the last few 100 Myr) [43]. In the Local Volume (< 10 Mpc), HST-resolution spectroscopy in the optical or UV will be essential to break $\alpha$-$M_u$ degeneracy, relying on relative line strength of different ionization stages (He I / He II, Si III / Si IV) and P-Cygni (C IV, N V) profiles to assign spectral type.

***Which SF quanta and processes underpin the low-density K-S relation?***
Observations,[44,45] theory[12,13,14,46] and simulations[47,48,49] suggest that the SFR/unit gas is dramatically reduced at low densities (Fig. 1), though stacking CO via an HI prior shows the linear relation between molecular column density and $\Sigma_{SFR}$ may be preserved.[50] However, recent results on this topic have relied on surface photometry in the UV, Hα and/or IR and are prone to systematic errors. A direct reassessment is needed of the resolved, low-density K-S relation[54,55,56] via counting of young massive star candidates (plus incorporation of any young clusters) occupying $N_{HI}$ bins. Relative motion of stars and gas is a complication, but can be controlled for by averaging over scales consistent with stellar lifetimes and typical velocity dispersions. Perhaps more importantly, this study would also yield a statistically meaningful description of the "SF products" [e.g. max/median cluster mass, max stellar mass] and small-scale clustering properties thereof, plus gas-based metrics (below), as a function of density and $\Sigma_{SFR}$. Such studies will provide direct constraints on the origin of scatter in the faint end of the K-S relation.

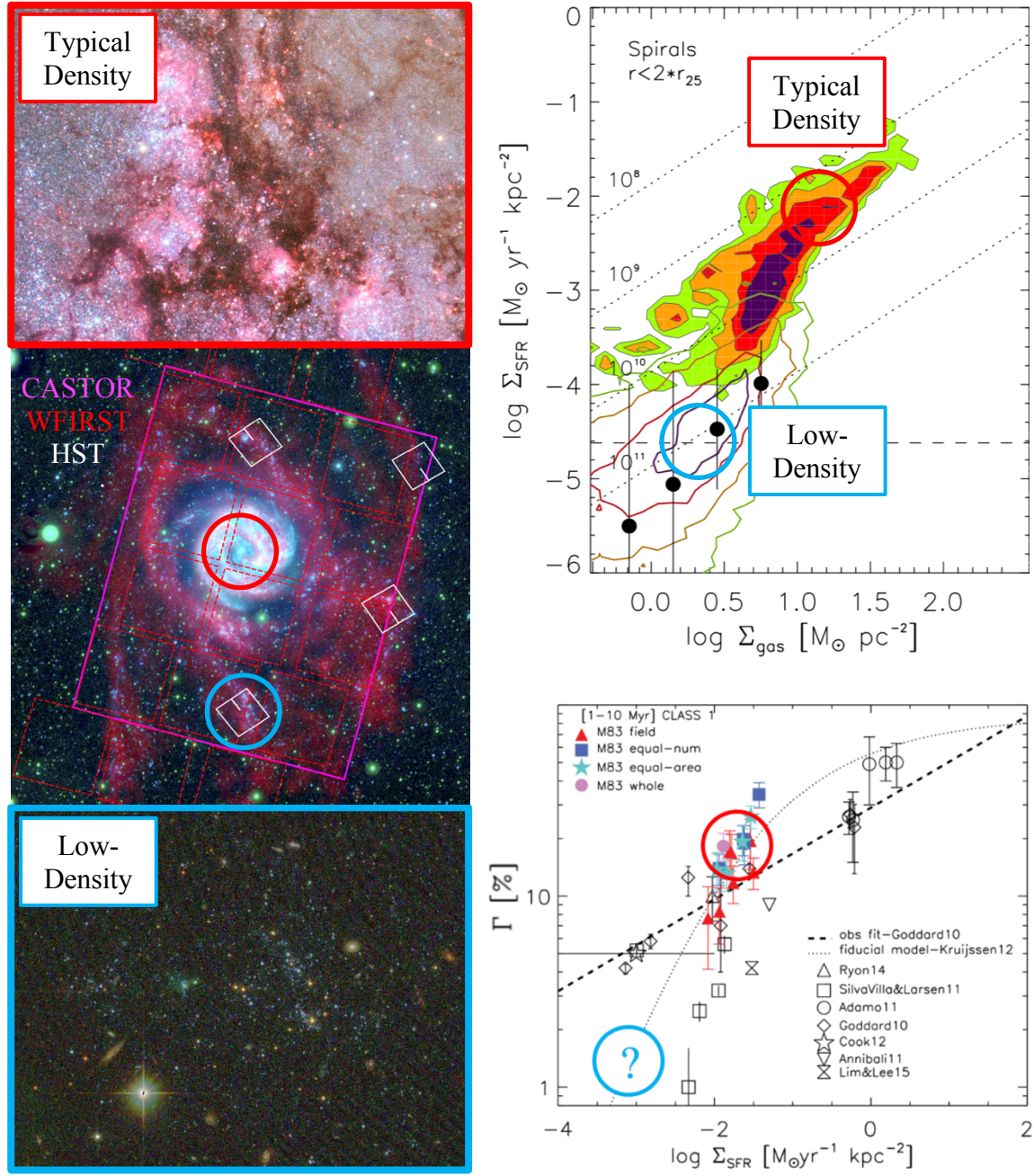

**Fig. 1:** How do the physics of star formation change as a function of density? Wide-field, high-resolution imaging is required to study SF in its primary quanta (stars, star clusters, gas clouds) over the complete range of densities probed by nearby galaxies, with new work needed on the low-density, HI-dominated regime. *(Left-center)* M83 (*d*=4.5 Mpc) and its archetypal extended UV disk. GALEX FUV/NUV is shown in (blue/green) with HI gas (red). A magenta rectangle shows the CASTOR FOV, dashed red squares show the WFIRST FOV and small white rectangles mark existing outer disk HST data with blue imaging. *(Left, top+bottom)* M83 HST imaging at typical and low-density, respectively. *(Right-top)* The K-S relation between SFR surface density $\Sigma_{SFR}$ and gas surface density, from Bigiel+2010[45]. *(Right-bottom)* The relation between the fraction of stars formed in bound clusters and $\Sigma_{SFR}$, from Adamo 2017[18]. *(In all panels)* Regions highlighted by red circles correspond to the well-studied central regions of galaxies while blue circles correspond to the low densities characteristic of galaxy outskirts, typical dwarf galaxies, tidal dwarfs, and LSB galaxies (including the sub-population of HI-bearing ultra-diffuse galaxies, UDGs[51], called HUDS[52]).

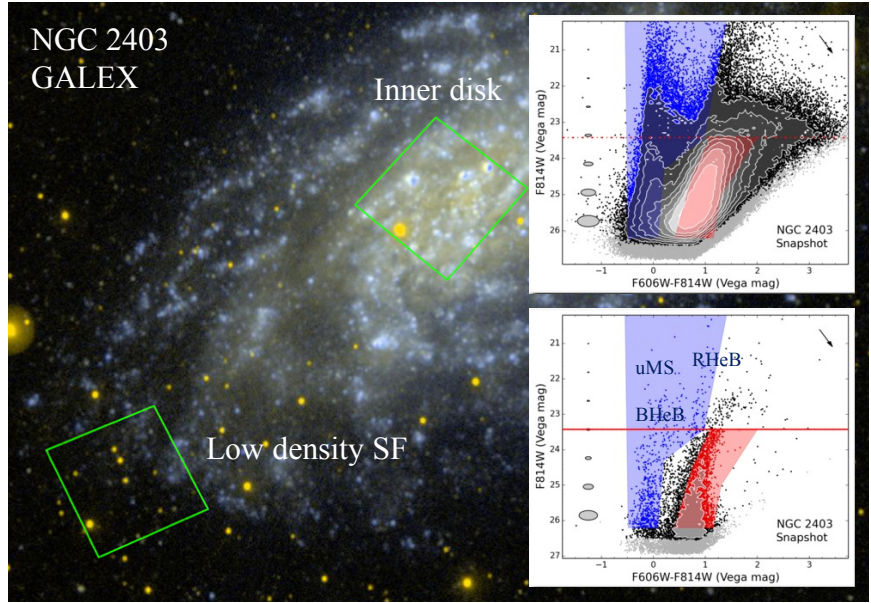

**Fig. 2:** CMDs of NGC2403 from GHOSTS[53] illustrating the difficulty of collecting a statistical sample of stars and clusters at low-density. The HST *field probing low-density shows only a few upper main sequence (uMS) stars* and a modest core He-burning (HeB) population, but is otherwise dominated by old stars. Few stellar clusters are found in this field. The inner disk has a well-populated uMS and many clusters. To accumulate a similar statistical sample in the LSB regime would be very costly using HST.

### *Diagnosing small scale physical conditions of the ISM at low-density*

The studies above will constrain "how" SF outcomes change at low-density. To explain "why," we need to diagnose cloud-scale (<50 pc) physical conditions in the ISM, then compare them to the resulting SF quanta. We must measure: (1) How prevalent is the cold neutral medium (CNM), the dense HI phase thought to be a precursor for SF, in these environments? (2) How abundant is dust? (3) What are the properties of star-forming atomic and molecular gas clouds? Precursor CNM clouds can be traced HI absorption[57], [CII] emission[58], or directly by high resolution HI spectral line observations (via detection of high brightness networks[59] or CNM/WNM spectral profile decomposition[60,61], possible with a factor 10 increase in radio observatory surface brightness sensitivity). Dust shielding, rather than molecular cooling is thought to be the key pre-condition for actual SF[17,48,49]. Outer disk dust abundance can be traced by sensitive far-IR mapping, important to measure at cloud scales. The immediate sites of SF are likely to be bound clouds composed of a mixture of atomic and molecular gas, but the properties of such faint clouds are essentially unexplored due to mapping speed limits even with ALMA. It has been extremely hard to identify fruitful targets for molecular line imaging[62,28] (perhaps because stellar feedback quickly disperses tenuous clouds). If likely sites of dense clumps can be located with improved HI imaging, their substructure, turbulence, chemical make-up and self-gravity are all of interest to deduce what sets the stellar output. Timescales for cloud formation/destruction can be probed by correlating cloud--SF site displacements versus age of the SF episode.[63,24,64] Statistical constraints on many questions can come by correlating average physical quantities (e.g., column[65] or line width distribution functions, cloud mass functions[66]) against young population indicators.

## Necessary Observing Capabilities and Present Outlook

This path forward *requires innovative capabilities: (1) UV-optical imaging with ~HST resolution and sensitivity but with extremely wide FOV (Table 1) to inventory stars and clusters at low $\Sigma_{SFR}$. (2) LUVOIR or HabEx for highly-multiplexed UV-optical spectral typing of O star candidates within ~10 Mpc. (3) cm-wave radio interferometry with 10× JVLA HI sensitivity to characterize dense HI gas clouds on small scales. (4) Efficient FIR spectroscopic mapping to capture dust properties and ISM cooling. (5) Deployment of heterodyne arrays on existing mm-wave telescopes.*

*(1)* CASTOR[67,68,69] (the Cosmological Advanced Survey Telescope for Optical and ultraviolet Research) is a Canadian initiative now in a Science Maturation Study phase. It provides 0.15" imaging simultaneously in *UV, U, g* to HST-like depths over a 0.56° x 0.44° field (5σ limits of 27.4, 27.4, 27.1 ABmag, respectively, in ~22 min/field primary survey exposure). If approved by the CSA, the CASTOR mission aims for a 2027 launch. CASTOR and WFIRST are synergistic as they have similar resolution and wide-fields, yet probe complementary spectral ranges.

*(2)* Both LUVOIR and HabEx mission concepts include micro-shutter-arrays that could be used for efficient spectroscopic (R>300) confirmation of individual massive stars in low-density SF regions. IMF work will benefit greatly from maximizing the mirror sizes to reach larger distances.

*(3)* A next-generation VLA (ngVLA)[70] as proposed would permit HI absorption work against very faint (more densely distributed) background sources and ~1" spectral line imaging to resolve dense atomic gas clouds which may host SF. ngVLA wide-field mapping of HI[29] in faint outer disks will be as transformative for small-scale atomic gas features as ALMA was for molecular clouds in inner disks of nearby galaxies. Both phases of the upcoming Square Kilometer Array (SKA) also offer a leap in HI capabilities, but their timelines, as well as US access, remain unclear.

*(4)* Origins Space Telescope FIR imaging spectroscopy, combined with HI maps, will constrain dust-to-gas ratio and interstellar radiation field at low-density and probe the [C II] cooling line.

*(5)* To survey wide areas for faint, sparse molecular line emission, large heterodyne arrays on single dish telescopes (e.g., GBT or LMT) or arrays (e.g. ALMA) can multiply the survey speed.

Table 1: Ultra-wide-field, HST-resolution UV-optical Imaging Capability

| Characteristic | Requirement | Specific Science Driver |
|---|---|---|
| Field of view | ≥ 0.2 square deg. (at least 100x HST/WFC3) | Efficiency to accumulate integrated SFR at low $\Sigma_{SFR}$ |
| Angular resolution | ≤ 0.15" FWHM (0.05"/pixel) Ideally diffraction limited at 1000 Å | Must resolve OB assoc. and differentiate individual stars from clusters at Local Volume edge, $d \sim 11$ Mpc |
| Spectral range | 2000-5000 Å. Ideally to 900 Å | To characterize young, hot populations and dust |
| Filters | Minimum (*NUV, U, B*) Ideally add *FUV*, plus narrow *NUV* bands on/beside 2175 Å absorption feature | Three bands for reddening-free *Q* or SED fits (with WFIRST), controlling for extinction |
| Sensitivity | In UV, equivalent to HST with F300X filter, e.g. 5σ limit NUV(AB)=27 in 45 min orbit | Efficiency to accumulate integrated SFR at low $\Sigma_{SFR}$ |
| Special characteristics | • Simultaneous imaging in multiple bands<br>• Grism spectroscopy | • Survey efficiency<br>• Sp. typing [age dating] of bright stars [clusters] |